\documentclass[aps,prl,reprint,showpacs,superscriptaddress]{revtex4-1}
\usepackage[english]{babel}
\usepackage{amsmath,amssymb,bbm,graphicx,color,comment,txfonts}
\usepackage{dsfont}
\usepackage[normalem]{ulem}
\usepackage{physics}
\usepackage[bookmarks=true,colorlinks,citecolor=blue,urlcolor=blue]{hyperref}

\begin{document}

\date{\today}

\title{Frustration-Free Control and Absorbing-State Transport in Entangled State Preparation}

\author{T. Dörstel}
\affiliation{Universität Innsbruck, Institut für Theoretische Physik, Technikerstraße 21a, 6020 Innsbruck, Austria}
\affiliation{Universit\"at zu K\"oln, Institut f\"ur Theoretische Physik, Zülpicher Str. 77a, 50937 Cologne, Germany}
\author{T. Iadecola}
\affiliation{Department of Physics, The Pennsylvania State University, University Park, PA 16802, USA}
\affiliation{Institute for Computational and Data Sciences, The Pennsylvania State University, University Park, PA 16802, USA}
\affiliation{Materials Research Institute, The Pennsylvania State University, University Park, Pennsylvania 16802, USA}
\author{J.~H. Wilson}
\affiliation{Department of Physics and Astronomy, Louisiana State University, Baton Rouge, LA 70803, USA}
\affiliation{Center for Computation and Technology, Louisiana State University, Baton Rouge, LA 70803, USA}
\author{M. Buchhold}
\affiliation{Universität Innsbruck, Institut für Theoretische Physik, Technikerstraße 21a, 6020 Innsbruck, Austria}
\affiliation{Universit\"at zu K\"oln, Institut f\"ur Theoretische Physik, Zülpicher Str. 77a, 50937 Cologne, Germany}

\begin{abstract}
We study frustration-free control, a measurement-feedback protocol for quantum state preparation that extends the concept of frustration-free Hamiltonians to stochastic dynamics. The protocol drives many-body systems into highly entangled target states, common dark states of all measurement projectors, through minimal local unitary corrections that realize an absorbing-state dynamics without post-selection. We show that relaxation to the target state is governed by emergent transport of nonlocal charges, such as singlet excitations in SU$(2)$-symmetric dynamics. While measurement-feedback annihilates compatible charge configurations, both measurement and scrambling unitaries induce charge transport and thus determine the convergence time. Mapping a baseline model of SU$(N)$ SWAP measurements with local corrections to a solvable absorbing random walk yields a runtime scaling $t \sim L^z$ with transport exponent $z=2$. Simulations of Motzkin and Fredkin chains reveal subdiffusive scaling $z \ge \tfrac{8}{3}$, confirming the transport picture and suggesting strategies for controlled entangled-state preparation and charge-transport probing in monitored quantum dynamics.
\end{abstract}

\maketitle

\emph{Introduction.}-- Engineering resourceful quantum many-body states is central to quantum information and quantum simulation. While Hamiltonian or unitary gate-based sequences remain a key paradigm~\cite{HVA2015,Zaletel2020,Causer2024}, measurement-based protocols with feedback offer robust, potentially advantageous schemes~\cite{GefenSteering2020,Morales2024,Langbehn2024,Piroli2021,Piroli2024,LuLessa2022}. Measurement-induced dynamics has a long tradition in quantum information  \cite{KITAEV20032,SurfaceCode,Gottesman_1999,Nielsen_2003,MBQC2001}, and has recently been propelled by advances in quantum platforms enabling mid-circuit measurements \cite{Bluvstein_2023, fossfeig2023, Baeumer2024}. Of particular interest are protocols realizing measurement-induced entanglement or control transitions~\cite{Szyniszewski2019, Skinner2019,LiFisher2019,LiFisher2018,Gullans_2020,Zabalo2020,Nahum2021,JianVasseur2020,Iadecola2023,Klocke2025,Lavasani2021}. Feedback schemes targeting dark or absorbing steady states~\cite{GauMajoranaDark,CooperAbsorbing2024,Sierant2023,DiPercIons,DiPercPiroli,RoyGefen2020} have demonstrated that deterministic control can be achieved by enforcing local constraints that render the desired state globally attractive.

\begin{figure}
    \centering
    \includegraphics[width=\columnwidth]{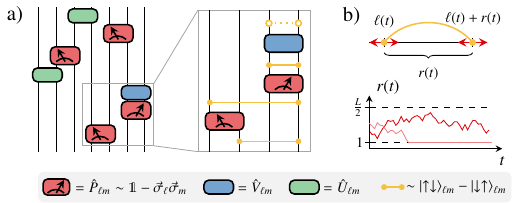}
    \caption{\textbf{Frustration-free control and absorbing random walks.}  
Schematic circuit that drives a spin chain into the entangled ground state manifold of the Heisenberg ferromagnet. Local measurements (red) compatible with the target leave the state unchanged, while incompatible outcomes trigger minimal unitary corrections (blue) that steer it toward the absorbing target state. Optional scrambling unitaries (green) preserve the target and accelerate convergence. (b) SU$(2)$-symmetric measurement and scrambling unitaries induce Brownian motion of singlet endpoints. Feedback annihilates short singlets of length $r(t)=1$. The diffusion-annihilation dynamics map to an absorbing random walk, governed by the transport exponent $z$.}
    \label{fig:Cartoon}
\end{figure}

In this work, we study a class of absorbing-state-inspired protocols designed to prepare target states with multipartite entanglement. The scheme extends the concept of frustration-free Hamiltonians to open-system dynamics~\cite{Puente2024,Gefen2024, YahuiLi2025}, and we refer to it as \emph{frustration-free control}. It employs a set of local projectors whose joint kernel defines a unique, entangled many-body state. The system is iteratively steered toward this state via repeated projective measurements and local unitary corrections conditioned on measurement outcomes. Crucially, the protocol implements an absorbing-state mechanism on individual quantum trajectories without post-selection.

Our goal is to uncover the physical mechanism underlying convergence: how local measurement and feedback operations build up global entanglement. We show that the approach to the target state is governed by an emergent absorbing-state dynamics in which nonlocal excitations, such as singlet pairs, diffuse and annihilate. This transport picture establishes the scaling of convergence times and provides a framework to optimize and generalize frustration-free control.

\emph{Frustration free control.}-- Consider a quantum system with $L$ degrees of freedom and a set of local projectors $\{\hat{P}_\alpha\}_{\alpha=1,...,L}$ satisfying $\hat{P}_\alpha^2 = \hat{P}_\alpha$, such that their common kernel uniquely defines a  many-body dark state  manifold $\{ |\psi_D\rangle\}$ via 
\begin{equation}
    \hat{P}_\alpha |\psi_D\rangle = 0 \quad \forall\, \alpha.
\end{equation}
Consider a circuit where projectors $\hat{P}_\alpha$ are measured in a random spatiotemporal pattern~\footnote{Structured patterns, such as Floquet circuits, may also be used, but the core strategy remains unchanged.} and feedback unitaries $\hat V_\alpha$ are applied, conditioned on the outcome. If the measurement yields $0$, the state is locally compatible with the target; if it yields $1$, a local unitary $\hat{V}_\alpha$ is applied, satisfying $[\hat{V}_\alpha, \hat{P}_\alpha] \neq 0$, to rotate the state towards the kernel of $\hat{P}_\alpha$. In addition, one can interleave unconditioned \emph{scrambling} unitaries $\hat U_\alpha$ that preserve the target manifold $\hat U_\alpha|\psi_D\rangle=|\psi_D\rangle$, while redistributing local excitations and accelerating relaxation. The circuit structure is illustrated in Fig.~\ref{fig:Cartoon}. 

When all projectors commute, measurements are independent and target a product state, reachable in $\mathcal{O}(L)$   operations. 
In contrast, non-commuting projectors $[\hat{P}_\alpha,\hat{P}_{\alpha'}]\!\ne\!0$ generate interference between measurements and feedback, turning control into a genuine many-body process that builds entanglement. 
This motivates the term \emph{frustration-free control}, in analogy with frustration-free Hamiltonians $\hat{H}=\sum_\alpha\hat{P}_\alpha$ whose common ground state minimizes competing local terms.  

For non-commuting measurements, deviations from the target state appear as \emph{non-local excitations}, such as singlets with well-defined endpoints [Fig.~\ref{fig:Cartoon}(b)]. Measurements and scrambling unitaries move these endpoints, while feedback annihilates adjacent pairs, leading to dynamics that map onto an \emph{absorbing random walk} with relaxation time set by the transport exponent~$z$. This mechanism, derived analytically for SU($N$)-symmetric SWAP measurements and verified numerically for Motzkin and Fredkin chains~\cite{Chirame2025}, provides a unifying description of convergence under feedback. As a concrete realization, we consider a spin-$\tfrac{1}{2}$ chain with SU$(2)$-symmetric measurements and SU$(2)$-symmetry-breaking feedback—previously studied numerically in Ref.~\cite{hauser2023continuous}—which admits an \emph{exact solution} via mapping to a random walker with an absorbing impurity. Extending this construction to include scrambling unitaries, arbitrary measurement geometries, imperfections, and SU($N$) symmetry yields a complete analytical framework for frustration-free control governed solely by transport.

\emph{Quantum master equation and order parameter.}-- The measurement-feedback protocol without scrambling defines a local quantum channel with trace-preserving Kraus maps:
\begin{equation}
    \hat{\rho} \longrightarrow \sum_{\sigma=0,1} \hat{K}_{\alpha,\sigma} \hat{\rho} \hat{K}_{\alpha,\sigma}^\dagger; \quad \hat{K}_{\alpha,0} = \mathds{1} - \hat{P}_\alpha, \quad \hat{K}_{\alpha,1} = \hat{V}_\alpha \hat{P}_\alpha.
\end{equation}
Let projectors $\hat P_\alpha$ be measured with probability $1-e^{-\gamma_\alpha t}$ in a time interval $t$. Taking the continuous-time limit $t\to dt$ and averaging over the stochastic application of measurement and feedback yields a Lindblad-type quantum master equation:
\begin{equation}\label{eq:QME}
    \partial_t \hat{\rho} = \sum_\alpha \sum_{\sigma=0,1} \gamma_\alpha \left( 
    \hat{K}_{\alpha,\sigma} \hat{\rho} \hat{K}_{\alpha,\sigma}^\dagger 
    - \tfrac{1}{2} \left\{ \hat{K}_{\alpha,\sigma}^\dagger \hat{K}_{\alpha,\sigma}, \hat{\rho} \right\} 
    \right)=\mathcal{L}\hat\rho,
\end{equation}
which drives the system toward the dark-state manifold $\{|\psi_D\rangle\}$. Each $|\psi_D\rangle$ is annihilated by all $\hat{K}_{\alpha,1}$ and left invariant by $\hat{K}_{\alpha,0}$. For the feedback $\hat{V}_\alpha$ defined above, this manifold forms the unique dynamically stable stationary solution of Eq.~\eqref{eq:QME}.

A useful metric to quantify the distance from the target state is the mean projector expectation value, $\mathds{E}(\langle \hat{P}_\alpha \rangle) \equiv \tfrac{1}{L} \sum_\alpha \langle \hat{P}_\alpha \rangle$, with $\langle \hat{P}_\alpha \rangle \equiv \text{tr}(\hat{\rho} \hat{P}_\alpha)$. By construction, $\langle \hat{P}_\alpha \rangle \geq 0$ for any physical state $\hat{\rho}$, and $\langle \hat{P}_\alpha \rangle = 0$ if and only if $\hat{P}_\alpha \hat{\rho} = 0$. Thus, $\mathds{E}(\langle \hat{P}_\alpha \rangle) = 0$ characterizes the global target state, while $\langle \hat{P}_\alpha \rangle$ serves as a local order parameter measuring the local deviation from the target. These local observables obey a closed evolution equation, derived from the master equation~\eqref{eq:QME}:
\begin{equation} \label{eq:OrderParQME}
    \partial_t \langle \hat{P}_\beta \rangle = \sum_{\alpha} \gamma_\alpha \left\langle \hat{P}_\alpha [\hat{V}_\alpha, \hat{P}_\beta] \hat{V}_\alpha^\dagger \hat{P}_\alpha - \tfrac{1}{2} [\hat{P}_\alpha, [\hat{P}_\alpha, \hat{P}_\beta]] \right\rangle.
\end{equation}
The double commutator in Eq.~\eqref{eq:OrderParQME} corresponds to dynamics without feedback ($\hat{V}_\alpha = \mathds{1}$). It induces measurement-induced dephasing, typically driving $\langle \hat{P}_\beta \rangle$ toward its value in the maximally mixed state. In contrast, the first term captures the effect of feedback and is responsible for steering the system toward the target state. This control is effective when the unitaries $\hat{V}_\alpha$ act \emph{off-diagonally} with respect to the projectors, such that $\hat{P}_\beta \hat{V}_\beta \hat{P}_\beta = \sqrt{(1-\lambda)} \hat{P}_\beta$ with $\lambda >0$. In this case, feedback induces a decay of $\langle \hat{P}_\beta \rangle$ with a rate proportional to $\gamma_\beta\lambda$.

\emph{Spin-$\tfrac12$ with SU$(2)$ symmetry.}-- To illustrate the feedback-control scheme introduced above, we consider a one-dimensional chain of $L$ spin-$\tfrac{1}{2}$ degrees of freedom subject to SU$(2)$-symmetric measurements and SU$(2)$-symmetry-breaking feedback. Specifically, the measurements project onto eigenstates of the SWAP operator $\mathcal{S}_{\ell,m}$, which exchanges the quantum states of spins $\ell$ and $m$. Since $\mathcal{S}_{\ell,m}^2 = \mathds{1}$, its eigenvalues are $\pm 1$, with the antisymmetric eigenvalue $-1$ corresponding to the maximally entangled \emph{singlet} state $|s_{\ell,m}\rangle=\tfrac{1}{\sqrt{2}}(\ket{\uparrow\downarrow}-\ket{\downarrow\uparrow})_{\ell,m}$. Using the Pauli vector $\vec{\sigma}_\ell = (\sigma^x_\ell, \sigma^y_\ell, \sigma^z_\ell)$ for spin $\ell$, the projector onto the singlet sector is $\hat{P}_{\ell,m} = \tfrac{1}{2} (\mathds{1} - \mathcal{S}_{\ell,m}) = \tfrac{1}{4} (\mathds{1} - \vec{\sigma}_\ell \cdot \vec{\sigma}_m)$. Each SWAP measurement conserves all total spin components $S^\nu_{\text{tot}} = \tfrac{1}{2} \sum_\ell \sigma^\nu_\ell$ for $\nu = x,y,z$, as well as the total angular momentum $\vec{J}^2 = \sum_\nu (S^\nu_{\text{tot}})^2 = \tfrac{1}{4}L(L+2) - \sum_{\ell \neq m} \hat{P}_{\ell,m}$, which quantifies the total number of singlets in the system. 

The target manifold is defined by $\hat{P}_{\ell,m}|\psi_D\rangle = 0$ for all $\ell, m$. It contains no singlets and exhibits maximal total spin $J = L/2$, implying that each bond $\hat P_{\ell,\ell+1}=0$ is symmetric. This subspace hosts $L+1$ fully symmetric Dicke states -- the ground states of the frustration free ferromagnetic Heisenberg model -- corresponding to different values of $S^z_{\text{tot}}$. To select a unique target, we fix the magnetization to $S^z_{\text{tot}} = 0$~\footnote{Note that we could equally well select the $W$-state by requiring $S^z_{\text{tot}} =\pm (L-1)$.}.

To reach the target state, we implement frustration-free control using nearest-neighbor measurements $\hat{P}_{\ell,\ell+1}$ along with corresponding local corrections $\hat{V}_{\ell,\ell+1}$. These unitaries are required to (i) act strictly locally on spins $\ell$ and $\ell+1$, making use of the measurement outcome, and (ii) break the $\mathrm{SU}(2)$ symmetry, i.e., $[\hat{P}_{\ell,\ell+1}, \hat{V}_{\ell,\ell+1}] \neq 0$, in order to rotate the system toward the target state. Single-spin Pauli rotations generated by $\sigma^\nu_{\ell}$ and $\sigma^\nu_{\ell+1}$ satisfy both requirements: they act locally and break $\mathrm{SU}(2)$ symmetry \footnote{Any two-spin unitary either preserves $\mathrm{SU}(2)$ or reduces to combinations involving single Pauli rotations and symmetric gates.}. To preserve $S^z_{\text{tot}}$, we choose, without loss of generality, the feedback unitary $\hat{V}_{\ell,\ell+1} = \sigma^z_\ell$.

\emph{Absorbing random walk.}-- In the absence of feedback, the singlet projectors \(\hat{P}_{\ell,m}\) are conserved under SU(2)-symmetric dynamics. They represent nonlocal excitations -- entangled Bell pairs -- anchored between sites $\ell$ and $m$. Their evolution follows Eq.~\eqref{eq:OrderParQME} with $\alpha=(\ell,\ell+1),\;\beta = (n,m)$. The double commutator yields
\begin{align}
   \sum_\ell \langle [\hat{P}_{\ell,\ell+1}, [\hat{P}_{n,m},\hat{P}_{\ell,\ell+1}]] \rangle = \tfrac{1}{2} \sum_\ell  D_{n,\ell} \langle \hat{P}_{\ell,m} \rangle + \langle \hat{P}_{n,\ell} \rangle D_{\ell,m},\nonumber
\end{align}
with \(D_{\ell,\ell'} = \delta_{\ell,\ell'+1} + \delta_{\ell, \ell'-1} - 2\delta_{\ell,\ell'}\) the lattice Laplacian. In the absence of control, $\mathrm{SU}(2)$ is intact and the expectation value $\langle \hat{P}_{n,m} \rangle$ evolves diffusively, with both endpoints of the singlet executing Brownian motion. The symmetry enforces conservation of the total singlet number and constrains the dynamics to local, diffusive rearrangements of singlet endpoints. 

Control explicitly breaks the $\mathrm{SU}(2)$ symmetry, opening a decay channel for singlets with minimal spatial extent,
\begin{align}
   \sum_\ell \left\langle \hat{P}_{\ell,\ell+1} [\hat{V}_{\ell,\ell+1},\hat{P}_{n,m}] \hat{V}_{\ell,\ell+1}^\dagger \hat{P}_{\ell,\ell+1} \right\rangle = -\lambda\, \langle \hat{P}_{n,m} \rangle \, \delta_{n,m\pm1},\nonumber
\end{align}
where $\lambda =1-  |\bra{s_{\ell,\ell+1}}\hat{V}_{\ell,\ell+1}\ket{s_{\ell,\ell+1}}|^2$ determines the decay [$\lambda=1$ for $\hat{V}_{\ell, \ell+1}=\sigma^z_\ell$]. This describes an annihilation process for singlets with endpoints on adjacent sites, which are rotated into the symmetric subspace.

To close the equations, we exploit translation invariance and label singlets by center-of-mass \(s = \tfrac{1}{2}(\ell + m)\) and relative distance \(r = |\ell - m| \bmod (L/2)\). The total weight of singlets of length \(r\), \(\mathcal{P}_r = \sum_\ell \langle \hat{P}_{\ell,\ell+r} \rangle\), evolves as (discrete Laplacian $D$)
\begin{align}
    \label{eq:rw_master}
  \partial_t \mathcal{P}_r = (D \mathcal{P})_r - 2\delta_{r,1} \, \mathcal{P}_r,
\end{align}
where we set $\gamma_\alpha\equiv1$. This describes a one-dimensional, classical random walk of the relative coordinate, with an absorbing impurity located at \(r = 1\). Hence, the total singlet weight \(\sum_r \mathcal{P}_r\) decays diffusively, with a global relaxation timescale that scales as \(L^z\), with the dynamical exponent \(z = 2\).

\begin{figure}
    \centering
    \includegraphics[width=\columnwidth]{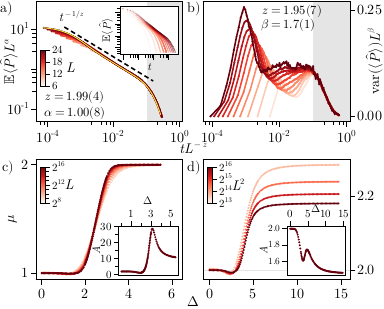}
    \caption{{\bf Frustration-free control of the Heisenberg chain.} 
(a) Exact simulations of the measurement-feedback protocol show a hydrodynamic regime with \(\mathds{E}(\langle \hat{P}_{\ell,\ell+1} \rangle) \sim t^{-1/z}\) and an asymptotic exponential decay \(\sim \exp(-t/L^z)\) for system sizes \(L \leq 24\), averaged over \(10^5\) trajectories (\(10^3\) for L=24). The yellow line is a single run of \(\mathcal{P}_1\) in Eq.~\eqref{eq:rw_master} for \(L=24\), the same initial state and no free parameters. 
(b) The variance of the order parameter exhibits strong fluctuations at the onset of the hydrodynamic regime (\(t \sim O(L^0)\)), plateaus during hydrodynamic evolution and is exponential cutoff at \(t \sim L^z\) (grey region). 
(c,d) Scaling exponents $\mu(L)$ of the cutoff time $t\sim L^{\mu(L)}$ as functions of system size and long-range exponent \(\Delta\) for absorbing random walks in one (c) and two (d) dimensions, obtained from solving Eq.~\eqref{eq:effME}. Inset: Total decay rate as a function of $\Delta$.
}
    \label{fig:su2_control}
\end{figure}

We numerically simulate the full quantum evolution of the mean \(\mathds{E}(\langle \hat{P}_\alpha \rangle)\) and the variance \(\mathrm{var}(\langle \hat{P}_\alpha \rangle) \equiv \frac{1}{\sum_\psi}\sum_\psi\langle\psi|\sum_\beta(\hat{P}_\beta - \mathds{E}(\langle \hat{P}_\alpha) \rangle)/L|\psi\rangle^2\) for system sizes up to \(L=24\), shown in Fig.~\ref{fig:su2_control}(a,b). After a short non-universal transient (\(t = \mathcal{O}(1)\)), the mean exhibits hydrodynamic scaling \(\mathds{E}(\langle \hat{P}_\alpha \rangle) \sim t^{-1/z}\), followed by an exponential finite-size cutoff \(\sim e^{-t/L^z}\) at \(t \sim L^z\), consistent with the absorbing random-walk picture at matching \(L\). The variance, taken over all simulated trajectories \(|\psi\rangle\) and not captured by the linear quantum master equation in Eq.~\eqref{eq:QME}, shows pronounced fluctuations at the onset of the hydrodynamic regime, saturates to a system-size–independent plateau, and then decays exponentially on the same characteristic timescale \(t \sim L^z\) as \(\mathds{E}(\langle \hat{P}_\alpha \rangle)\), indicating the absence of strong trajectory fluctuations.

\emph{Effect of unitary scrambling.}-- 
We now consider the impact of additional, unconditioned \emph{scrambling} unitaries $\hat U_\ell$, applied randomly in space and time with uniform rate $\gamma\kappa$. 
The gates are chosen from the set of SU$(2)$-symmetric unitaries of the form $\hat U_{\ell} = \exp(i\varphi\,\mathcal{S}_{\ell,\ell+1}) = \cos(\varphi) + i\sin(\varphi)\,\mathcal{S}_{\ell,\ell+1}$,\footnote{Any SU$(2)$ gate can be decomposed into a sequence of such local SWAP operations. Averaging over random realizations is therefore equivalent to using this elementary set.} and averaging each update step $\hat\rho \!\to\! \hat U_\ell \hat\rho \hat U_\ell^\dagger$ uniformly over $\varphi \!\in\! [0,2\pi]$ yields $\hat\rho \!\to\! \hat\rho - \tfrac{1}{4}[\mathcal{S}_{\ell,\ell+1},[\mathcal{S}_{\ell,\ell+1},\hat\rho]]$. 
With this process acting uniformly in space, the singlet dynamics in Eq.~\eqref{eq:rw_master} is modified by a renormalization of the diffusion constant, $D \!\to\! (1+\kappa)D$, i.e., scrambling enhances Brownian motion. 
Tuning~$\kappa$ rescales the eigenvalues of the effective diffusion operator $(1+\kappa)D$, accelerating convergence. For example, local scrambling at rate $\kappa \sim L^{\alpha}$ shifts the bare $L^{-2}$ level spacing of $D$ to $L^{\alpha-2}$ for $\kappa D$, reducing relaxation times to $t \sim \max{(L^{2-\alpha},1)}$.
Fast scrambling hence removes the diffusion bottleneck and turns a diffusion-limited absorbing process into annihilation-limited.

\emph{Generalization to arbitrary geometries and SU$(N)$.}--  The control scheme naturally extends to different lattice geometries of the spins. Consider $L^d$ spins in $d$ dimensions at positions \(\vec{\ell}, \vec{m} \in \mathds{R}^d\), with \(\hat{P}_{\vec{\ell},\vec{m}} = \tfrac{1}{2}(\mathds{1} - \mathcal{S}_{\vec{\ell},\vec{m}})\) projecting onto the antisymmetric state. Measuring \(\hat{P}_{\vec{\ell},\vec{m}}\) and applying local, magnetization-preserving feedback unitaries on one spin of each measured pair drives the system toward the fully symmetric subspace \(\hat{P}_{\vec{\ell},\vec{m}}|\psi\rangle = 0\), independent of geometry and the measurement sequence. The SU$(2)$ symmetry of the measurements, conserving the total spin and magnetization, remains intact. While this leaves the target state unchanged, the \emph{approach} to it -- i.e., the dynamical efficiency of the protocol -- depends on the geometry and measurement sequence.

We consider a translationally invariant setup where pairs \((\vec{\ell}, \vec{\ell} + \vec{r})\) are measured with rate \(\gamma_{|\vec{r}|}\). Since the projector algebra of SU$(2)$ SWAPs is independent of geometry, the previous analysis applies. Defining the 'diffusion kernel' \(\tilde{D}_{\vec{r},\vec{s}} = \gamma_{|\vec{r} - \vec{s}|}\), the total weight of singlets of distance $\vec r$ follows [c.f.~\cite{endmatter}]
\begin{align}\label{eq:effME}
  \partial_t \mathcal{P}_{\vec r} = \sum_{\vec s\neq 0} \left( \tilde{D}_{\vec r,\vec s} - 2\gamma_{|\vec r|} \delta_{\vec{r},\vec{s}} \right) \mathcal{P}_{\vec s} = -\sum_{\vec s\neq0} \tilde{H}_{\vec r,\vec s} \mathcal{P}_{\vec s}.
\end{align}
Probability conservation requires $\sum_{\vec{s}\neq0}\tilde{D}_{\vec{r},\vec{s}}=0$ for all $\vec {r}$, i.e., $\sum_{\vec r\neq0}\gamma_{|\vec r|}=-\gamma_0$. We normalize the total rate such that $\gamma_0=1$, which implies an extensive number of $L^d dt$ measurements per time step $dt$. The asymptotic decay of the total weight is governed by the lowest eigenvalue \(\tau > 0\) of the operator \(\tilde{H}\), i.e., \(\sum_{\vec r} \mathcal{P}_{\vec r} \sim \exp(-\tau t)\).

We consider an algebraic decay of measurement rates, \(\gamma_{|\vec{r}|} =\Gamma_{\Delta} |\vec{r}|^{-\Delta}\) for $|\vec{r}|>0$, with the range exponent \(\Delta\) interpolating between nearest-neighbor (\(\Delta \to \infty\)) and all-to-all (\(\Delta \to 0\)) measurements and $\Gamma_{\Delta}>0$ fixing the normalization for any finite $L<\infty$. Our goal is to determine the scaling of the decay rate \(\tau(\Delta, L)=A(\Delta)\cdot L^{-\mu(\Delta)}\) with the total number of spins \(L^d\). The limiting case \(\Delta \to 0\) is exactly solvable~\cite{endmatter}, yielding \(\tau =2 L^{-d}\) and \(\mu(\Delta \to 0) = d\). 

We analyze square lattices in \(d = 1\), Fig.~\ref{fig:su2_control}(c), and \(d = 2\), Fig.~\ref{fig:su2_control}(d). For each \(\Delta\), we numerically extract the smallest eigenvalue \(\tau(\Delta, L)\) of \(\tilde{H}\), and compute the exponent $\mu(\Delta)$ via the beta function $
\mu(\Delta) \equiv \frac{\log[\tau(\Delta, 2L)] - \log[\tau(\Delta, L)]}{\log(2)}$. For \(d = 1\), a sharp transition at a critical decay exponent \(\Delta_c \approx 2.1\) separates the infinite-range regime \((\mu = 1)\) from the short-range diffusive regime \((\mu = 2)\). In $d=2$, $\mu$ displays a narrow range \(\mu \in [2, 2.3]\) and  crossing points at \(\mu = d\), such that the relaxation is governed by the infinite-range scaling.

\begin{figure}
    \centering
    \includegraphics[width=\columnwidth]{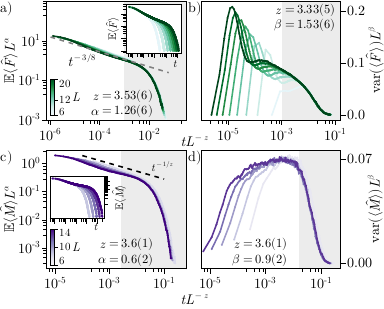}
    \caption{\textbf{Anomalous transport in frustration-free control.} Measurement-feedback dynamics for (a,b) the spin-\(\tfrac{1}{2}\) Fredkin and (c,d) spin-1 Motzkin chains exhibit an intermediate regime \(\mathds{E}(\langle \hat{Q}_\alpha \rangle)\!\sim\! t^{-1/z}\) followed by exponential decay \(\mathds{E}(\langle \hat{Q}_\alpha \rangle)\!\sim\!\exp(-t/L^{z})\), signaling subdiffusive transport with \(z>2\).  
For the Motzkin chain, data up to \(L\!\le\!14\) yield \(z\!\approx\!3.6\). The Fredkin chain shows a hydrodynamic regime \(\mathds{E}(\langle \hat{P}_\alpha \rangle)\!\sim\! t^{-z}\) with \(z\!=\!\tfrac{8}{3}\), crossing over to an asymptotic decay consistent with \(z\!\gtrsim\!3.3\)~\cite{Fradkin}.
}
    \label{fig:AnomalousTransport}
\end{figure}

For spin-$(N-1)/2$ states, the algebra of SWAPs \(\mathcal{S}_{\ell,m} = \sum_{\alpha,\beta=1}^{N} |\alpha\beta\rangle\langle\beta\alpha|_{\ell,m}\) is independent of $N$. Hence, the control scheme readily generalizes to SU$(N)$: the SWAP projector dynamics, Eq.~\eqref{eq:effME} and all subsequent analysis apply equally to general $N$. We stress that for $N>2$, the symmetric subspace, i.e., the control target space, is strictly larger than the subspace of maximal total angular momentum~\cite{endmatter}.

\emph{Fredkin and Motzkin chains.}-- We illustrate frustration-free control in kinetically constrained models, namely the Fredkin and Motzkin chains~\cite{MainFredkinPaper}. Both are frustration-free spin systems with unique ground states exhibiting multipartite entanglement. Although their excitation sectors are not separated by a non-Abelian symmetry, they are particularly interesting because (i) their excitations, nonlocal charge configurations represented by \emph{Motzkin words}, are balanced spin configurations with two well-defined endpoints, and (ii) their transport dynamics is subdiffusive, characterized by a dynamical exponent $z \ge \tfrac{8}{3}$~\cite{vasseurfredkin1,vasseurfredkin2}.

Frustration-free control is implemented as above: we measure local projectors $\hat Q_\ell$ and apply feedback operations steering the system toward the target state $|\psi\rangle$ with $\hat Q_\ell |\psi\rangle = 0$. For the Fredkin dynamics, we consider a spin-$\tfrac{1}{2}$ chain with $\hat Q_\ell \equiv \hat F_\ell = (1 - \sigma^z_{\ell-1}) \hat P_{\ell,\ell+1} + \hat P_{\ell-1,\ell} (1 + \sigma^z_{\ell+1})$, where $\hat P_{\ell,\ell+1}$ projects onto the singlet state of the two spins. For the Motzkin chain, we consider $L$ spin-1 degrees of freedom and $\hat Q_\ell \equiv \hat M_\ell = |U\rangle\langle U|_{\ell,\ell+1} + |D\rangle\langle D|_{\ell,\ell+1} + |F\rangle\langle F|_{\ell,\ell+1}$, with $|U\rangle_{\ell,\ell+1} = (\ket{\uparrow0} - \ket{0\uparrow})/\sqrt{2}$, $|D\rangle_{\ell,\ell+1} = (\ket{\downarrow0} - \ket{0\downarrow}/\sqrt{2}$, and $|F\rangle_{\ell,\ell+1} = (\ket{\uparrow\downarrow} - |00\rangle)/\sqrt{2}$. Upon positive measurement outcomes, local feedback is applied as $V_\ell = \exp(i \tfrac{\pi}{2} \hat Z_\ell)$. Starting from a zero-magnetization state, the protocol drives the system into a unique steady state.

Numerical simulations of the order parameter $\mathds{E}(\langle \hat Q_\ell \rangle)$ and its variance for both the Motzkin and Fredkin chains reveal a scaling collapse consistent with an absorbing random walk, see Fig.~\ref{fig:AnomalousTransport}. As in the SWAP case, we identify an intermediate transport regime $O(L) < t < L^z$, where $\mathds{E}(\langle \hat Q_\ell \rangle) \sim t^{-1/z}$, followed by an exponential cutoff $\mathds{E}(\langle \hat Q_\ell \rangle) \sim \exp(-t L^{-z})$. The Fredkin chain exhibits transport consistent with $z = \tfrac{8}{3}$ at intermediate times, while the scaling collapse of the exponential tail suggests $z \simeq 3.3$--$3.5$, indicating a strong suppression of transport at long times. For the Motzkin chain, we find consistently large exponents, $z \approx 3.6$. While previous studies~\cite{Fradkin,vasseurfredkin1} reported $z = \tfrac{8}{3}$ for the lowest-lying excitation of the Hamiltonian $\hat H = \sum_\ell \hat Q_\ell$, values $z > 3$ have been observed for higher excitations~\cite{Fradkin}, which appear to govern the relaxation dynamics relevant for feedback control.

\emph{Imperfect control.}-- Realistic settings are subject to imperfections such as noise or measurement errors. To understand the effect of detrimental noise, interfering with the absorbing state condition, we consider an elementary phenomenological model of reversible defect dynamics 
within the absorbing random walk framework: let random defects occur with rate $\eta$ and create (annihilate) a singlet of length $|\vec{r}|$ with combinatorial probability $2/L$ ($4\mathcal{P}_{\vec r}$). This modifies Eq.~\eqref{eq:effME}:
\begin{align}\label{eq:effDEF}
  \partial_t \mathcal{P}_{\vec r} = 2\eta - \sum_{\vec s} \left( \tilde{H}_{\vec r,\vec s} + 4\eta\,\delta_{\vec r, \vec s} \right) \mathcal{P}_{\vec s}.
\end{align}
The stationary state is set by a balance between decay and noise. It is maximally mixed  $\mathcal{P}_{\vec{r}} = \tfrac{1}{2}$ if $\eta\gg \tilde H$, while it scales as $\mathcal{P}_{\vec r} \sim \eta\, L^{z}$ for $\eta\ll\tilde H$ and hence is governed by the dynamical exponent $z$. This result is confirmed by exact numerical simulations of imperfect measurements, see~\cite{endmatter}. To maintain high-fidelity, either $z$ needs to be small, or errors must be suppressed with system size as $\eta \sim L^{-z}$. This quantifies the robustness of frustration free control and further allows to probe $z$ under realistic imperfections.

\emph{Outlook.}-- Frustration-free control prepares entangled states through measurement–feedback–driven annihilation of nonlocal charges. The dynamics is governed by charge transport, characterized by a global dynamical exponent $z$ that sets the runtime and efficiency. Remarkably, scrambling unitaries compatible with measurement constraints, such as symmetries, do not hinder but instead accelerate convergence. This observation enables optimization via enhanced transport, for example through long-range measurements, unitary dynamics, or environmental coupling, and offers an experimental route to probe the dynamical exponent $z$.

Future directions include extensions to open systems, implementations in atomic and solid-state platforms, and control schemes targeting nonsymmetric or topologically ordered states. It is also natural to explore steering into frustration-free subspaces under chaotic dynamics, such as in quantum many-body scars~\cite{Shiraishi2017,LindbladScarA} or fermionic models~\cite{Frustrationfreefermion}, where absorbing-state or entanglement transitions may arise.

\begin{acknowledgments}
{\it Acknowledments.}--    T.~D. and M.~B. acknowledge support from the Deutsche Forschungsgemeinschaft (DFG, German Research Foundation) under Germany’s Excellence Strategy Cluster of Excellence Matter and Light for Quantum Computing (ML4Q) EXC 2004/1 390534769, and by the DFG Collaborative Research Center (CRC) 183 Project No.~277101999 - projects B01. M.B. acknowledges support from the Heisenberg programme of the Deutsche Forschungsgemeinschaft (DFG, German Research Foundation), project no. 549109008. T.I. and J.H.W. acknowledge support from the National Science Foundation under Grant Number DMR-2143635 (T.I.) and DMR-2238895 (J.H.W.). This work was initiated at Aspen Center for Physics, which is supported by National Science Foundation grant PHY-2210452.
    The numerical simulations have been performed on the RAMSES cluster at RRZK, University of Cologne.\\
\end{acknowledgments}

{\it Data availability}.-- 
All code for generating the figures  and the underlying numerical data is available on Zenodo~\cite{zenodo_FrustFreeControl}.

\bibliography{references}
\newpage
\phantom{Text}
\newpage

\appendix

\section{Appendix}

\emph{Absorbing random walk derivation}.-- 
For projectors \(\hat P_{\ell,m} = \tfrac{1}{2}(\mathds{1} - \mathcal{S}_{\ell,m})\) defined via the spin-\(j\) SWAP operator \(\mathcal{S}_{\ell,m}=\sum_{\alpha\beta}|\alpha\beta\rangle\langle\beta\alpha|_{\ell,m}\), the double commutator \([\hat P_{\ell,m}, [\hat P_{\ell,m}, \hat P_{s,n}]]\) vanishes if either \(\{\ell,m\}=\{s,n\}\) or all indices are distinct. It is nontrivial only when one index pair coincides and the other does not. Using \(\mathcal{S}_{\ell,m}^2 = \mathds{1}\), \(\mathcal{S}_{\ell,m}\mathcal{S}_{\ell,n}\mathcal{S}_{\ell,m} = \mathcal{S}_{m,n}\), and \( [\hat P_{\ell,m},[\hat P_{\ell,m},\hat P_{\ell,n}]] = -\tfrac{1}{8}[\mathcal{S}_{\ell,m},[\mathcal{S}_{\ell,m},\mathcal{S}_{\ell,n}]]\), one finds
\begin{align}
    [\hat P_{\ell,m},[\hat P_{\ell,m},\hat P_{\ell,n}]] &= -\tfrac{1}{4}(\mathcal{S}_{\ell,n} - \mathcal{S}_{m,n}) 
    = \tfrac{1}{2}(\hat P_{\ell,n}-\hat P_{m,n}).
\end{align}
This result is independent of the spatial dimension \(d\) and spin \(j\). Introducing the relative coordinate \(r=\mod(m-\ell,L/2)\) and the center of mass (COM) coordinate $s=(\ell+m)/2$ via \(\hat P_{\ell,m} = \hat P_{s-r/2, s+r/2} \equiv \hat P_{s,r}\) and collecting all nonvanishing terms in Eq.~\eqref{eq:OrderParQME} of the main text, one obtains
\begin{align}
    \partial_t \hat P_{s,r} &= \tfrac{1}{2} \sum_{r'\neq0} \gamma_{|r'|}\Big(-2\hat P_{s,r} (2-\delta_{r,-r'} - \delta_{r,r'})\nonumber\\
    &+\hat P_{s-r',r+r'} + \hat P_{s,r-r'}+\hat P_{s,r+r'} + \hat P_{s+r',r-r'}\Big)\nonumber\\
    & \vphantom{\Bigg(}-\lambda (\gamma_{|r|} \hat P_{s,r} + \gamma_{|-r|} \hat P_{s+r, -r}).
\end{align}
Summing both sides over \(s\) and introducing \(\mathcal{P}_r = \sum_{s} \langle\hat P_{s,r}\rangle\), the center of mass coordinate $s$ is eliminated due to translational invariance. This yields the equation
\begin{equation}
    \partial_t \mathcal{P}_r = \sum_{r'\neq0}\gamma_{|r'|}\big( \mathcal{P}_{r+r'} + \mathcal{P}_{r - r'} - \mathcal{P}_r (2 - \delta_{r,-r'} - \delta_{r,r'}\big) - 2\lambda \gamma_{|r|} \mathcal{P}_r,
\end{equation}
which due to $\gamma_{|r'|}=\gamma_{|-r'|}$ yields Eq.~\eqref{eq:effME} in the main text. 
The contribution from the feedback $\hat{V}_{\ell,\ell'}$ reduces to a factor $\lambda$, as 
\begin{equation}
    \hat{P}_{\alpha} V_\alpha P_\alpha V_\alpha ^\dagger P_\alpha = P_\alpha |\expval{V_\alpha}{s_\alpha}|^2 =P_\alpha |\tfrac{1}{4}\tr(V_\alpha)|^2.
\end{equation}
With this relation, the full contribution from the correction reduces to 
\begin{equation}
    \hat{P}_\alpha [\hat V_\alpha, \hat{P}_\beta]\hat V_\alpha ^\dagger \hat{P}_\alpha = \delta_{\alpha,\beta} (1 - |\tfrac{1}{4}\tr(V_\alpha)|^2 \hat{P}_\alpha) \equiv \delta_{\alpha,\beta} \lambda \hat{P}_\alpha,
\end{equation}
where the $\delta_{\alpha,\beta}$ is a result of $\hat{P}_{\ell,\ell'} \sigma^i_\ell \hat{P}_{\ell,\ell'} = 0$ for $i = x,y,z$. Studying single Pauli corrections that yield $\lambda = 1$ is sufficient: any two-site unitary can be represented by $e^{i\phi_\nu\sigma^\nu_\ell\sigma^\nu_{\ell'}}$ multiplied with single-Pauli operations, whereas $|\expval{e^{i\phi_\nu\sigma^\nu_\ell\sigma^\nu_{\ell'}}}{s_{\ell,\ell'}}|^2 = 1$. 

If we choose $\gamma_r = r^{-\Delta}$ with $\Delta \rightarrow \infty$, Eq.~\ref{eq:effME} is recovered. In the all-to-all measurement limit $\Delta = 0$, the equation can simplified even further. The equal weight sum over all terms yields $\sum_{r'} \mathcal{P}_{r+r'} = \sum_{r'} \mathcal{P}_{r-r'}$ and thus 
\begin{equation}
    \partial_t \mathcal{P}_{r} = \tfrac{2}{L^d - 1} \Big(\sum_{r'\neq0} \mathcal{P}_{r'} - (L^d-1)\mathcal{P}_r\Big).
\end{equation}
A matrix with this structure can be diagonalized directly, where the eigenvalue with the largest absolute value reads $\tau = -2 (L^d-1)^{-1}$. These limits confirm Fig.~\ref{fig:su2_control} with $\mu(\Delta = 0) = d$ and $A(\Delta = 0) = 2$.

\emph{SU\((2)\) charge dynamics}.-- 
The Hilbert space of two spin-\(\tfrac{1}{2}\) particles is spanned by the singlet and triplet states: \(\ket{s}=\tfrac12(\ket{\uparrow\downarrow}-\ket{\downarrow\uparrow})\), \(\ket{t_1}=\ket{\uparrow\uparrow}\), \(\ket{t_0}=\tfrac12(\ket{\uparrow\downarrow}+\ket{\downarrow\uparrow})\), and \(\ket{t_{-1}}=\ket{\downarrow\downarrow}\). The operator \(\hat P_{\ell,m}=\tfrac{1}{2}(\mathds{1}-\mathcal{S}_{\ell,m})\) projects onto the singlet: \(\hat P_{\ell,m}\ket{s}_{\ell,m}=\ket{s}_{\ell,m}\). Enforcing \(\hat P_{\ell,\ell+1}\ket{\psi}=0\) for all \(\ell\) implies that every bond—and therefore every pair \((\ell,m)\)—is in a superposition of triplets. For a chain of length \(L\), the \(L+1\) states satisfying this condition are the permutation-symmetric Dicke states \(\ket*{D_L^k}\), which are symmetric superpositions of \(k\) spins up and \(L-k\) spins down.

Any measurement \(\hat P_{\ell,m}\) commutes with the total angular momentum \(\vec J^2\) (see main text). Hence, the total number of singlets is conserved under measurement, and only the singlet endpoints are reshuffled. For example, for \(\ket{\psi}=\ket{\sigma}_1\ket{s}_{2,3}\), measuring a singlet on \((1,2)\) yields \(\hat P_{1,2}\ket{\psi}\sim \ket{s}_{1,2}\ket{\sigma}_3\), effectively swapping spins \(1\leftrightarrow3\). A triplet measurement gives \((\mathds{1}-\hat P_{1,2})\ket{\psi}\sim \ket{\psi}-\ket{\sigma}_2\ket{s}_{1,3}\). The correction \(\hat V_{\ell,\ell+1}=\sigma^z_\ell\) rotates a singlet into a triplet, \(\sigma^z_\ell\ket{s}_{\ell,m}=\ket{t_0}_{\ell,m}\), and thus, when applied immediately after a measurement \(\hat P_{\ell,m}\), removes exactly one length-1 singlet.

\emph{Symmetric subspaces in SU$(N)$.}-- 
Unlike the SU$(2)$ spin-\(\tfrac{1}{2}\) case, the maximally symmetric subspaces of SU$(N)$ are not identical to the maximal-\(J\) subspaces, although the latter are contained within the former. Consider spin-1 (SU$(3)$): two spins decompose into singlet, triplet, and quintet sectors with \(\vec{J}^2=0,1,2\), respectively. Despite having minimal total angular momentum, the singlet \(\ket{s}=\tfrac{1}{\sqrt{3}}(\ket{1,-1}+\ket{-1,1}-\ket{0,0})\) is symmetric under particle exchange. For \(j>\tfrac{1}{2}\), the maximal-\(J\) sector is therefore embedded in (but not identical to) the symmetric subspace. While measurements of SU$(N)$ SWAP operators (with \(N=2j+1\)) project naturally into the symmetric subspace, isolating the maximal-\(J\) sector requires more elaborate measurement schemes and generally lacks a closed-form time evolution.

\emph{Motzkin words.}-- 
The Fredkin operator \(\hat{F}_{\ell} = (1 - \sigma^z_{\ell-1}) \hat{P}_{\ell,\ell+1} + \hat{P}_{\ell-1,\ell} (1 + \sigma^z_{\ell+1})\) implements a similar logic to the \(\operatorname{CSWAP}\) (Fredkin) gate. It can be rewritten as 
\begin{equation*}
    F_{\ell} = \tfrac{1}{2}(1-\mathcal{CS}_{\ell,\ell+1,\ell+2}) + \tfrac{1}{2}(1-\sigma^x_{\ell+2}\mathcal{CS}_{\ell+2, \ell+1, \ell} \sigma^x_{\ell+2}),
\end{equation*}
where $\mathcal{CS}_{i,j,k}$ is the CSWAP gate that swaps sites $j$ and $k$, if $i$ is in the $\ket{\uparrow}$ state.
These dynamics preserve the total magnetization \(S^z\) and a family of nonlocal charges known as Motzkin (or Dyck) words. Spin configurations are mapped to Motzkin paths by identifying spin-up and spin-down with ascending and descending steps, \(\downarrow\to/\) and \(\uparrow\to\backslash\), and spin-zero with a flat step, \(0\to-\). A configuration thus becomes a height profile, e.g., \(\ket{\uparrow\downarrow\uparrow\uparrow\downarrow\downarrow}\to/\backslash//\backslash\backslash\). Any configuration forming a closed profile, \( / \ldots \backslash\), is a Motzkin word. Fredkin gates do not mix Motzkin with non-Motzkin words; they only shift the endpoints of matched parentheses. On three qubits, the Fredkin gate acting on computational-basis states has two outcomes: if no Motzkin word is present (e.g., \(///\), \(\backslash\backslash\backslash\), \(\backslash//\), or \(\backslash\backslash/\)), the action vanishes; otherwise, it moves the Motzkin word and produces an antisymmetric superposition of the shifted and input states.

In the Motzkin model, Motzkin words are extended to include flat steps; otherwise, their interpretation and dynamics remain equivalent. See Refs.~\cite{DellAnna2019,pronko2025periodicmotzkinchainground} for details on both chains and on Motzkin words.

\begin{figure}
    \centering
    \includegraphics[width=\linewidth]{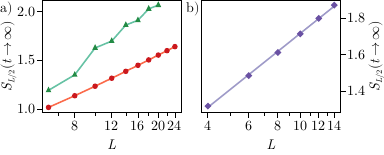}
    \caption{\textbf{Stationary-state half-chain entanglement.} Numerical half-chain entanglement in the \(S_z=0\) sectors versus analytical expressions. (a) The spin-\(\tfrac{1}{2}\) SU$(2)$ model (circles) coincides with the entanglement of the Dicke state \(S_{L/2}(\ket*{D_L^{L/2}})\) (lower solid line). The Fredkin model (triangles) converges to \(\ket{A}-\ket*{D_L^{L/2}}\) (upper solid line). (b) Consistent with analytical results for the open Motzkin chain \cite{DellAnna2019}, the stationary state of the periodic Motzkin chain exhibits logarithmic entanglement scaling (solid line: linear fit on a semi-log plot).}
    \label{fig:ent-scaling}
\end{figure}

\emph{Target states and entanglement scaling.}-- 
The target states of the SU$(N)$, Fredkin, and Motzkin measurement-feedback protocols are the common kernels of \(\{\hat P_\alpha\}\), \(\{\hat F_\alpha\}\), and \(\{\hat M_\alpha\}\), respectively. Each protocol conserves the total magnetization \(S_z=\sum_\ell \hat Z_\ell\). To select a unique target, we work in the \(S_z=0\) subspace, where the targets exhibit nontrivial multipartite entanglement.

In the SU$(2)$ model, the target is the symmetric Dicke state \(\ket*{D_L^{L/2}}\). For a Dicke state \(\ket*{D_L^k}\), the entanglement entropy of a block of size \(\ell\) is~\cite{DickeStatesEntropy}
\begin{equation}
    S_{\ell} (\ket*{D_{L}^{k}}) = -\binom{L}{k}^{-1} \sum_{i=0}^{\min(\ell, k)} \binom{\ell}{i} \binom{L - \ell}{k - i} \ln \left[ \binom{L}{k}^{-1} \binom{\ell}{i} \binom{L - \ell}{k - i} \right].\nonumber
\end{equation}
We confirm logarithmic entanglement growth for the target state; see Fig.~\ref{fig:ent-scaling}(a).

For the Fredkin circuit at \(S_z=0\), two states satisfy \(\hat F_{\ell}|\psi\rangle=0\): the Dicke state \(\ket*{D_L^{L/2}}\) and the anomalous ground state \(\ket{A}\)~\cite{MainFredkinPaper},
\begin{equation}
    \ket{A} =\binom{L}{n}^{-1/2}\sum_{\omega \in \text{comb}(L)} (-1)^{m(\omega)}\ket{\omega},
\end{equation}
where \(\text{comb}(L)\) denotes the unique permutations with \(n=L/2\) up and \(n\) down spins, and \(m(\omega)\) is the absolute maximum height of configuration \(\omega\). Figure~\ref{fig:ent-scaling}(a) shows that the controlled Fredkin circuit’s stationary state is the antisymmetric superposition \(\ket{\psi} = \ket{A} - \ket*{D_L^{L/2}}\). The simulated entanglement entropy matches that of \(\ket{\psi}\). The oscillatory increase likely reflects the enhanced number of Dyck paths for system sizes divisible by four, as also seen in open-boundary Fredkin chains~\cite{DellAnna2019}. 

For the Motzkin circuit, the target is the ground state of the periodic Motzkin chain \(H_{\text{Motzkin}, \text{pbc}} = \sum_i M_{i,i+1} + M_{L,1}\), conjectured to be a generalized spin-1 Dicke state~\cite{pronko2025periodicmotzkinchainground}. We also find logarithmic entanglement scaling with system size; see Fig.~\ref{fig:ent-scaling}(b).

\begin{figure}[t]
    \centering
    \includegraphics[width=\columnwidth]{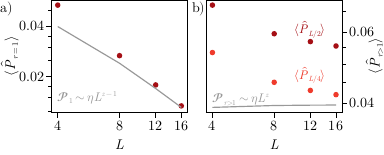}
    \caption{\textbf{Imperfect control.} Scaling of the stationary singlet expectation \(\mathcal{P}_r\) with noise rate \(\eta = L^{-2}\) versus exact simulations. The gray lines show \(M^{-1}\vec{\eta}\). (a) Length-1 singlets scale as \(\mathcal{P}_1 \sim \eta L^{z-1}\); (b) longer singlets obey \(\mathcal{P}_r \sim \eta L^z\) for \(r>1\). Dots display the stationary singlet expectations for the corresponding \(r\), exhibiting the same scaling.}
    \label{fig:noise}
\end{figure}

\emph{Confirming the imperfect control picture.}-- 
To validate the phenomenological noise model of Eq.~\eqref{eq:effDEF}, we simulate a circuit with measurement imperfections in the SU$(2)$ setup of the main text. We assume an apparatus that occasionally misreports outcomes. To implement a defect rate \(\eta\) following the phenomenological model, we take singlet outcomes to be falsely flagged—interpreted as triplets and not corrected—with probability \(p_{s,\text{false}} = \tfrac{1}{2}(1 + e^{-\eta})\). Conversely, triplet outcomes are (incorrectly) corrected with probability \(p_{t,\text{false}} = \tfrac{1}{2}(1 - e^{-\eta})\). The simulations reproduce the predicted scaling of \(\vec{\mathcal{P}}(t\!\to\!\infty)\); see Fig.~\ref{fig:noise}. Singlets of length \(r=1\) scale as \(\mathcal{P}_1 \sim \eta L^{z-1}\), while longer ones scale as \(\mathcal{P}_r \sim \eta L^{z}\), yielding an average scaling \(\sim L^{z}\) for the total singlet number.

\emph{Scaling of the diffusion operator versus decay rate in 1D.}-- 
The spectrum of the translationally invariant diffusion operator \(D_{\vec r}\) follows from the Fourier transform of the algebraic kernel \(D_{\vec r}\sim |\vec r|^{-\Delta}\). In momentum space,
\begin{align}
    D_{q}\sim 
    \begin{cases}
        q^2, & \text{for } \Delta>3,\\
        q^2 \log(q), & \text{for } \Delta=3,\\
        q^{\Delta-1}, & \text{for } 1<\Delta<3.
    \end{cases}
\end{align}
For a finite system of size \(L\), momentum modes have spacing \(\delta q\sim 1/L\), so the excitation gap follows directly. For a localized absorbing impurity at \(\vec r=1\), this gap sets the decay rate. However, when the dispersion becomes long-ranged, the impurity also acquires algebraic spatial support. In the infinite-range limit, the decay rate scales as \(\sim 1/L\), independent of the diffusion-operator dispersion. Thus, as the impurity range increases, the dynamics cross over from diffusion-limited to decay-limited absorbing random walks. At the transition in Fig.~\ref{fig:su2_control}, the system therefore switches from diffusion- to decay-limited behavior, rather than from short- to long-range diffusion.

\end{document}